\documentclass[preprintnumbers,amsmath,amssymb]{revtex4}

\usepackage{epsfig}
\usepackage{graphics}
\usepackage{graphicx}% Include figure files
\usepackage{dcolumn}% Align table columns on decimal point
\usepackage{bm}% bold math
\newcommand{\vect}[1]{\mathbf{#1}}

\begin{document}

\title{Emergence of a common generalized synchronization manifold in network motifs of structurally different time-delay systems}% Force line breaks with \\
\author{R.~Suresh$^a$}
\author{D.~V.~Senthilkumar$^{b}$}
\email{skumarusnld@gmail.com}
\author{M.~Lakshmanan$^c$}%\footnote{Author for correspondence}
\author{J.~Kurths$^{d,e,f}$}

\affiliation{
$^a$Centre for Nonlinear Science \& Engineering, School of Electrical \& Electronics Engineering, SASTRA University, Thanjavur 613 402, India.\\
$^b$School of Physics, Indian Institute of Science Education and Research, Thiruvanandapuram 695016, India\\
$^b$Centre for Nonlinear Dynamics, School of Physics, Bharathidasan University, Tiruchirapalli 620 024, India\\
$^c$Potsdam Institute for Climate Impact Research, 14473 Potsdam, Germany\\
$^d$Institute of Physics, Humboldt University, 12489 Berlin, Germany\\
$^e$Institute for Complex Systems and Mathematical Biology, University of Aberdeen, Aberdeen AB24 3UE, United Kingdom}

\begin{abstract}
We point out the existence of a transition from partial to global generalized synchronization (GS) in symmetrically coupled structurally different time-delay systems of different orders using the auxiliary system approach and the mutual false nearest neighbor method. The present authors have recently reported that there exists a common GS manifold even in an ensemble of structurally nonidentical scalar time-delay systems with different fractal dimensions and shown that GS occurs simultaneously with phase synchronization (PS). In this paper we confirm that the above result is not confined just to scalar one-dimensional time-delay systems alone but there exists a similar type of transition even in the case of time-delay systems with different orders. We calculate the maximal transverse Lyapunov exponent to evaluate the asymptotic stability of the complete synchronization manifold of each of the main and the corresponding auxiliary systems, which in turn ensures the stability of the GS manifold between the main systems. Further we estimate the correlation coefficient and the correlation of probability of recurrence to establish the relation between GS and PS. We also calculate the mutual false nearest neighbor parameter which doubly confirms the occurrence of the global GS manifold.
\end{abstract}

%\pacs{05.45.Xt,05.45.Pq,02.30.ks,89.75.-k}% PACS, the Physics and Astronomy
\maketitle

\section{\label{sec1}Introduction}
During the past couple of decades, the phenomenon of chaos synchronization has been extensively studied in coupled nonlinear dynamical systems from both theoretical and application perspectives due to its significant implications in diverse natural and man-made systems \cite{pikovsky01,lakshmanan10}. In particular, various types of synchronization,  including complete synchronization (CS) where
the coupled systems evolve identically, phase synchronization (PS) referring to the entrainment in the phase
of the interacting systems while their amplitude remains uncorrelated, and generalized synchronization (GS) where there
exist some functional relation between the coupled systems, etc., have been identified. All these types of synchronization have been investigated mainly in identical systems and in systems with some parameter mismatch. Very occasionally, it has been studied in distinctly nonidentical (structurally different) systems. But in reality, structurally different systems are predominant in nature and engineering and very often the phenomenon of GS is responsible for their evolutionary mechanism and proper functioning  of such structurally different systems. Typical examples include the cooperative functions of brain, heart, liver, lungs, limbs, etc., synchronization in living systems, coherent coordination of different parts of machines, synchronization between cardiovascular and respiratory systems \cite{schafer98}, different populations of species \cite{blasius99,amritkar06}, in epidemics \cite{grenfell00,earn98}, in visual and motor systems \cite{farmer98,sebe06}, in climatology \cite{stein11,maraun05}, in paced maternal breathing on fetal \cite{leeuwen09}, etc. Further, it has also been shown that GS is more likely to occur in spatially extended systems and complex networks (even in networks with identical nodes, due to the large heterogeneity in their nodal dynamics) \cite{shang09,hung08,moskalenko12}. In addition, GS has been experimentally observed in laser systems \cite{uchida03}, liquid crystal spatial light modulators \cite{rogers04}, microwave electronic systems \cite{dmitriev09} and has applications in secure communication devices \cite{murali98,moskalenko10}. Therefore understanding the evolutionary mechanisms of many natural systems necessitates the understanding of the underlying intricacies involved in the GS phenomenon. %In all such interacting systems there exists a synchronization with a functional relationship (an important condition for GS) between the interactions. Therefore, understanding the collective synchronized evolution of distinctly nonidentical systems is crucial and challenging.

GS has been well studied and understood in unidirectionally coupled systems \cite{basnarkov14,kocarev96,rulkov95,abarbanel96,brown98,pyragas96}, but still it remains largely unexplored in mutually coupled systems. Only a limited number of studies have been carried out on GS in mutually coupled systems even with parameter mismatches \cite{Koronovskii14,zheng02,hung08,guan09,chen09,hu10,guan10,shang09,moskalenko12,shahverdiev09,acharyya13} and rarely in structurally different dynamical systems with different fractal dimensions \cite{ouannas15,boccaletti00}. Recent investigations have revealed that GS emerges even in symmetrically (mutually) coupled network motifs made of identical systems, and that it also plays a vital role in achieving coherent behavior of the entire network \cite{sorino12,moskalenko12}. As almost all natural networks are heterogeneous in nature, the notion of GS has been shown to play a vital role in their evolutionary mechanisms \cite{pikovsky01}. Therefore to unravel the role of GS in the evolution of such a large networks, it is crucial to understand the underlying dynamics involved in the onset and emergence of GS in heterogeneous network motifs composed of structurally different systems. It is also to be noted that the notion of PS has been widely investigated in mutually coupled essentially different (low-dimensional) chaotic systems \cite{pikovsky01}, while the notion of GS in such systems has been largely ignored. 

The relation between PS and GS have been reported in low dimensional systems \cite{parlitz96,zheng00,stankovski14} and it has been shown that in general GS always leads to PS in unidirectionally coupled chaotic systems. In contrast, PS may occur in cases where the coupled systems show no GS \cite{parlitz96} attributing to the stronger nature of  GS. Further, Zhang and Hu \cite{zheng00} have demonstrated that GS is not necessarily stronger than PS, and in some cases PS comes after GS with increasing coupling strength depending upon the degree of parameter mismatch. They have concluded that PS (GS) emerges first for low (high) degree of parameter mismatch and that they both occur simultaneously for a critical range of mismatch in low-dimensional systems \cite{zheng00}. An attempt to unify the concepts of PS an GS has also been made in ref \cite{stankovski14}. In addition, the transition from PS to GS as a function of the coupling strength has been demonstrated in coupled time-delay systems with parameter mismatch \cite{senthil07}.   Despite these clear understanding on GS and PS transition in unidirectionally coupled systems, to the best of our knowledge, the relation between GS and PS in mutually coupled systems
has not yet been investigated so far. In general, the notion of GS and its relation with PS in mutually coupled systems, particularly in structurally different systems with different fractal dimensions including time-delay systems, need much deeper understanding which remains as a void in the literature.

In line with the above discussion, we have reported briefly the existence of GS in symmetrically coupled networks of structurally different scalar one-dimensional time-delay systems using the auxiliary system approach \cite{senthil13}. In this paper, we will extend our investigations to non-scalar, higher dimensional heterogeneous time-delay systems to examine whether GS can still persist between strongly heterogeneous systems (with different orders) and to understand the underlying dynamical transitions. In particular, in this paper we will demonstrate the emergence of a transition from partial to global GS in mutually coupled structurally different time-delay systems with different fractal (Kaplan-Yorke) dimensions and most importantly in systems with different orders using the auxiliary system approach and the mutual false nearest neighbor (MFNN) method.  Here the term partial GS refers to the state where only a few of the coupled
systems are entrained to the common GS mainfold, whereas the term global GS refers to the state where all the coupled systems are in
GS. In addition, we have also provided a detailed explanation about structurally different time-delay systems with different fractal dimensions and on the attracting GS manifold.
%(we use MFNN method only in array configuration with $N=2,~3$ and $4$ systems, in other configurations due to complexity in coupling it is not possible to obtain the reconstructed attractor with similar time-scales in structurally different time-delay systems).
We use the Mackey-Glass (MG) \cite{mackey77}, a piecewise linear (PWL) \cite{senthil06,senthil07}, a threshold piecewise linear (TPWL) \cite{suresh13} and the Ikeda time-delay \cite{ikeda80} systems to construct strongly heterogeneous network motifs. The main reason to consider time-delay systems in this study is that even with a single time-delay system, one has the flexibility of choosing systems with different fractal dimensions just by adjusting their intrinsic delay alone, which is a quite attracting feature of time-delay systems from the modelling point of view \cite{appeltant11}. Further, time-delay occurred within the systems are ubiquitous in several real situations. In particular, intrinsic time-delay can be found in neuronal models, where neurons are connected to autapse (a self-synapse or a specialized connection between a neuron and itself), which can be described by time-delayed feedback in closed loop \cite{qin14,qin14a,song15}. In addition, intrinsic time-delay can also observed in ecology, epidemics, physiology, physics, economics, engineering and control systems, \cite{lakshmanan10} which inevitably require delay for a complete description of the dynamical system (note that an intrinsic delay is different from connection delays which arise between different systems due to finite signal propagation time).  Propagation delay induced synchronization in different types of networks has also been studied in detail in the literature \cite{wang11, wang11a, wang11b, wang10, wang10a, wang09}.

In particular, we report that there exists a common GS manifold even in structurally different time-delay systems. In other words, there exists a functional relationship even for systems with different fractal dimensions, which maps them to a common GS manifold. Further, we also wish to emphasize that our results are not confined to just scalar one-dimensional time-delay systems alone but we confirm that there exists a similar type of synchronization transition even in the case of time-delay systems of different orders. Particularly, we demonstrate that the phenomenon of GS manifests in a system of Ikeda time-delay system (first order time-delay system) mutually coupled with a Hopfield neural network (a second order time-delay system), and in a system of a MG time-delay system (first order time-delay system) mutually coupled with a plankton model (a third order system with multiple delays) to establish the generic nature of our results. Stability of GS manifold in unidirectionally coupled systems is usually determined by examining the conditional Lyapunov exponents of the synchronization manifold~\cite{kocarev96,pyragas96} or the Lyapunov exponents of the coupled system itself~\cite{moskalenko12}. Here, we will estimate the maximal transverse Lyapunov exponent (MTLE) to determine the asymptotic stability of the CS manifold of each of the systems with their corresponding auxiliary systems starting from different initial conditions, which in turn asserts the stability of GS between the original structurally different time-delay systems. Further, we will also estimate the cross correlation (CC) and the correlation of probability of recurrence (CPR) to establish the relation between GS and PS (where GS and PS always occur simultaneously in structurally different time-delay systems). CC essentially gives a much better statistical average of the synchronization error, which is being widely studied to characterize CS. Further, CPR is a recurrence quantification tool~\cite{marwan07}, which effectively characterizes the existence of PS especially in highly non-phase-coherent hyperchaotic attractors usually exhibited by time-delay systems~\cite{senthil06}. It is also to be noted that the auxiliary system approach has some practical limitations. This method fails for systems whose dynamical equations are not known and
%and even though the dynamical equations are known, the auxiliary response system can be designed only with finite accuracy (especially in experimental simulations one cannot have an exact replica of the original response system).
also CS between response and auxiliary systems arises only when their initial conditions are set to be in the same basin of attraction. Due to the above limitations of the auxiliary system approach, we have also calculated the MFNN which doubly confirms our results. %In addition, an analytical stability condition using the Krasovskii-Lyapunov theory is also deduced in suitable cases.

The remaining paper is organized as follows: In Sec.~\ref{sec2}, we will describe briefly the notion of structurally different time-delay systems with different fractal dimensions with examples. In Secs.~\ref{sec3} we briefly describe the mathematical formulation of the auxiliary system approach for mutually coupled structurally different time-delay systems and in Sec.~\ref{sec3a} and \ref{sec4}, we will demonstrate the existence of a transition from partial to global GS in $N=2$ mutually coupled time-delay systems using the auxiliary system approach and the MFNN method, respectively. Further, we will consider an array of $N=4$ mutually coupled time-delay systems and discuss the occurrence of partial and global GS transition in Sec.~\ref{sec6}. The above synchronization transition in time-delay systems with different orders is demonstrated in Sec.~\ref{sec7} and finally we summarize our results in Sec.~\ref{sec8}.
\begin{table}[h]
	{\begin{tabular}{l l}\\[-2pt]
			\toprule\\
			CS &~~~~Complete Synchronization\\
			GS &~~~~Generalized Synchronization\\
			PS &~~~~Phase Synchronization\\
			MFNN &~~~~Mutual False Nearest Neighbor\\
			MG &~~~~Mackey-Glass \\%[1pt]
			PWL &~~~~Piecewise Linear\\%[2pt]
			TPWL &~~~~Threshold Piecewise Linear \\%[2pt]
			MTLE &~~~~Maximal Transverse Lyapunov Exponent \\%[1pt]
			LE &~~~~Lyapunov Exponent\\
			CC &~~~~Cross Correlation\\
			CRR &~~~~Correlation of Probability of Recurrence\\
			\botrule
		\end{tabular}}
		\caption{A list of abbreviations used in the paper.}
		\label{table2}
	\end{table}
	
\section{\label{sec2}Structurally Different Time-delay Systems with Different Fractal Dimensions}
In this section, we consider structurally different first order scalar time-delay systems with different fractal dimensions. Here, structurally different time-delay systems refer to systems exhibiting chaotic/hyperchaotic attractors with different phase space geometry characterized  by different degrees of complexity. Despite the similarity in the structure of their evolution equations, the nature of chaotic attractors, the number of their positive LEs and their magnitudes characterizing the rate of divergence and the degree of complexity as measured by the Kaplan-Yorke dimension ($D_{KY}$) of the underlying dynamics are different even for the same value of time-delay because of the difference in the nonlinear functional form.

As an illustration, first let us consider a symmetrically coupled arbitrary network of structurally different scalar time-delay systems. Then the dynamics of the $i$th node in the network is represented as
\begin{equation}
\dot{\vect{x}}_i= -\alpha_i \vect{x}_i(t)+\beta_i \vect{f}_{i}(\vect{x}_i(t-\tau_{i}))
-\varepsilon \sum_{j=1}^N G_{ij} \vect{x}_{j},
\label{eqn1}
\end{equation}
where $i=1,...,N$, and $N$ is the number of nodes in the network, $\alpha_i$'s and $\beta_i$'s are system's parameters, $\tau_i$'s are the time-delays, the smooth continuous function of the $i$th node is defined as $\vect{f}_{i}(\vect{x}_{i})$, $\varepsilon$ is the overall coupling strength and $G$ is a Laplacian matrix which determines the topology of the arbitrary network. Here we have studied a linear array with open end boundary conditions. For the MG time-delay system, we choose the nonlinear function \cite{mackey77}
\begin{equation}
f_{1}(x(t-\tau_{1})) = \frac{x_{1}(t-\tau_{1})}{(1+(x_{1}(t-\tau_{1})^{10}))},
\label{eqn1a}
\end{equation}
and for the PWL system the nonlinear function is given as \cite{senthil06,senthil07}
\begin{eqnarray}
f_{2}(x)=
\left\{
\begin{array}{cc}
0, &  x \leq -4/3  \\
            -1.5x-2,&  -4/3 < x \leq -0.8 \\
           x,&  -0.8 < x \leq 0.8 \\
            -1.5x+2,&  -0.8 < x \leq 4/3 \\
0, &  x > 4/3.
         \end{array} \right.
\label{eqn1b}
\end{eqnarray}
For the TPWL system we choose the form of the nonlinear function as given by \cite{suresh13}
\begin{equation}
f_{3}(x) = AF^{*}-Bx,
\label{eqn1c}
\end{equation}
with
\begin{eqnarray}
F^{*}=
\left\{
\begin{array}{cc}
-x^{*},&  x < -x^{*}  \\
            x,&  -x^{*} \leq x \leq x^{*} \\
            x^{*},&  x > x^{*}, \\
         \end{array} \right.
\label{eqn1d}
\end{eqnarray}
and for the Ikeda time-delay system the nonlinear function is given by \cite{ikeda80}
\begin{equation}
f_{4}(x(t-\tau_{4}))=sin(x(t-\tau_{4})).
\label{eqn1e}
\end{equation}
\begin{figure}
\centering
\includegraphics[width=0.8\columnwidth]{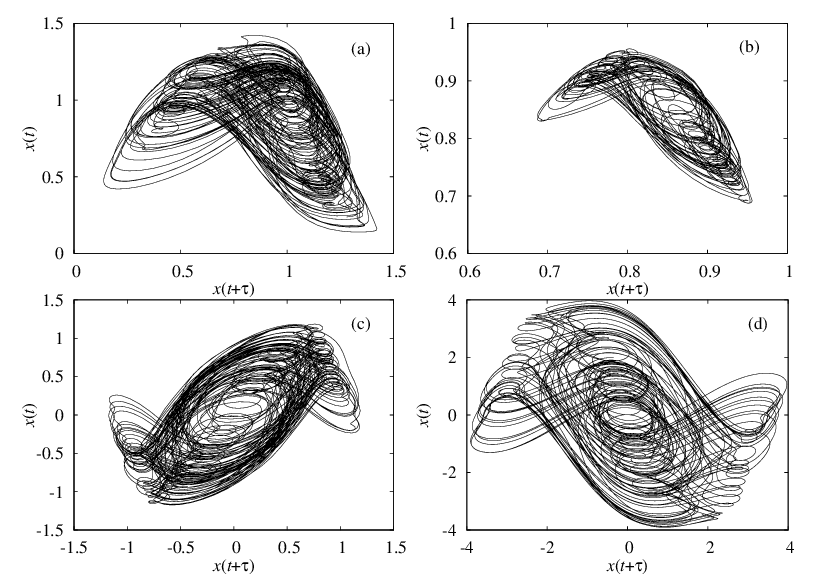}
\caption{\label{fig1}Hyperchaotic attractors of (a) Mackey-Glass, (b) piecewise linear, (c) threshold piecewise linear and (d) Ikeda time-delay systems for the choice of the parameters given in Table~\ref{table1}.}
\end{figure}

The parameter values for the above time-delay systems are chosen throughout the paper as follows: \\
1) For the MG systems: We choose $\beta_{1}=0.5$, $\alpha_{1}=1.0$, and $\tau_{1}=8.5$;\\
2) For the PWL systems: We choose $\beta_{2}=1.0$, $\alpha_{2}=1.2$, $\tau_{2}=10.0$, $p_{1}=0.8$ and $p_{2}=1.33$;\\
3) The parameter values for the TPWL are fixed as $\beta_{3}=1.0$, $\alpha_{3}=1.2$, $\tau_{3}=7.0$, $A=5.2$, $B=3.5$ and $x^{*}=0.7$ and \\
4) for the Ikeda time-delay system we choose $\beta_{4}=1.0$, $\alpha_{4}=5.0$, $\tau_{4}=7.0$. 

\begin{figure}
\centering
\includegraphics[width=0.8\columnwidth]{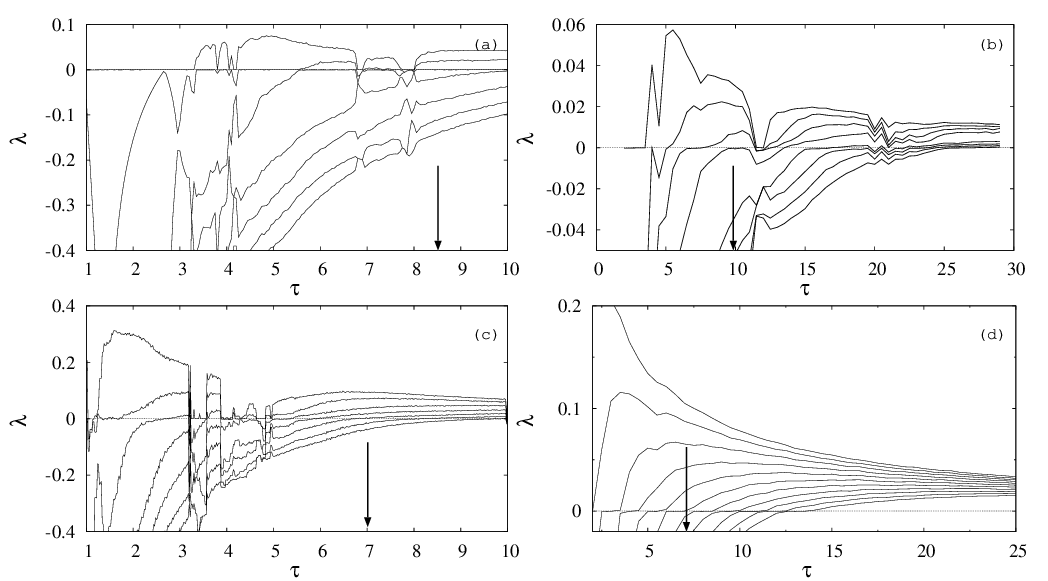}
\caption{\label{fig2}First few largest Lyapunov exponents of (a) Mackey-Glass, (b) piecewise linear, (c) threshold piecewise linear and (d) Ikeda time-delay systems, as a function of the time-delay $\tau$. Arrows point the value of the time-delay we have considered in our analysis. The choice of the corresponding parameter values of all the four systems are given in Table~\ref{table1}.}
\end{figure}

The hyperchaotic attractors of the uncoupled MG, PWL, TPWL and Ikeda time-delay systems are depicted in Figs.~\ref{fig1}(a)-(d), respectively. The first few largest LEs of all the above four (uncoupled) time-delay systems are shown as a function of the time-delay $\tau$ in Figs.~\ref{fig2}(a)-(d). It is clear from this figure that the number of positive LEs, and hence the complexity and dimension of the state space, generally increase with the time-delay. Further, the degree of complexity, measured by their number of positive LEs of the dynamics (attractors) exhibited by all the four systems are distinctly different even for the same value of time-delay. In fact, we have taken different values of delay for each of the systems, which are indicated by the arrows in Fig.~\ref{fig2}, to demonstrate the existence of suitable smooth transformation that maps the strongly distinct individual systems to a common GS manifold.

\begin{table}[h]
{\begin{tabular}{c c c c c c c}\\[-2pt]
\toprule
No ~~& System ~~~&\multicolumn{3}{c} {Choice of parameters} ~~&No. of Positive LEs ~~& $D_{KY}$ \\[6pt]
%\hline\\[-2pt]
& & $\beta_{i}$  ~~~~& $\alpha_{i}$ & $\tau_{i}$  &  & \\
\hline
1 ~~& MG ~~~& 0.5 ~~~~& 1.0 & 8.5 ~~& 2 ~~& 2.957 \\%[1pt]
2 ~~& PWL ~~~& 1.0 ~~~~& 1.2 & 10.0 ~~& 3 ~~& 4.414\\%[2pt]
3 ~~& TPWL ~~~& 1.0 ~~~~& 1.2 & 7.0 ~~& 4 ~~& 8.211\\%[2pt]
4 ~~& Ikeda ~~~& 1.0 ~~~~& 5.0 & 7.0 ~~& 5 ~~& 10.116\\%[1pt]
\botrule
\end{tabular}}
\caption{The parameter values, number of positive LEs and Kaplan-Yorke dimension ($D_{KY}$) of the structurally different time-delay systems.}
\label{table1}
\end{table}

We wish to emphasize especially the structural difference, as measured by their degree of complexity, between the hyperchaotic attractors of different scalar first order time-delay systems (Fig.~\ref{fig1}) which we have employed in this paper, are detailed in Table ~\ref{table1}: 1) the MG system has two positive LEs with $D_{KY}=2.957$ for $\tau_1=8.5$, 2) the PWL time-delay system has three positive LEs with $D_{KY}=4.414$ for $\tau_2=10.0$, 3) the TPWL time-delay system has four positive LEs with $D_{KY}=8.211$ for $\tau_3=7.0$ and 4) the Ikeda system has five positive LEs with $D_{KY}=10.116$ for $\tau_4=7.0$. The above facts clearly indicate that the real state space dimension explored by the flow of a time-delay system, which is essentially infinite-dimensional in nature, and the associated degree of complexity are characterized by the form of nonlinearity and the value of the time-delay irrespective of the similarity of the underlying evolution equations of the scalar first order time-delay systems.
\section{\label{sec3}Transition from Partial to Global GS in Mutually Coupled Time-delay Systems: Auxiliary System Approach}
It has already been known that the functional relationship between two different systems in GS is generally difficult to identify analytically. However GS in such systems can be characterized numerically by using various approaches, namely the mutual false nearest neighbor method \cite{rulkov95}, the statistical modeling approach \cite{schumacher12}, the phase tube approach \cite{koronvskii11}, the auxiliary system approach \cite{abarbanel96}, etc. Among all these methods the auxiliary system approach is extensively used to detect the presence of GS in unidirectionally coupled systems (both in numerical and experimental studies due to its simple and powerful implementation). Abarbanel et al. \cite{abarbanel96} first introduced this approach to characterize and confirm GS in dynamical systems  (when the system equations are known). The mathematical formulation of this concept was put forward by Kocarev and Parlitz \cite{kocarev96} for a drive-response configuration in low-dimensional systems. The formulation is based on the asymptotic convergence of the response and its auxiliary systems which are identically coupled to the drive system, starting from two different initial conditions from the same basin of attraction. The asymptotic convergence indeed ensures the existence of an attracting synchronization manifold (CS manifold between the response and auxiliary systems and GS manifold between the drive and response systems) \cite{kocarev96}. In other words, GS between the drive $\vect{x}$ and the response $\vect{y}$ systems occur only when the response system is asymptotically stable, that is $\forall~\vect{y_i}(0)~\&~\vect{x}(0)$ in the basin of the synchronization manifold one requires $\lim_{t \to \infty}\vert\vert \vect{y}(t,\vect{x}(0),~\vect{y}_1(0)) -\vect{y}(t,\vect{x}(0),~\vect{y}_2(0))\vert\vert=0$.

Now, we will extend this approach to a network for mutually coupled systems. For simplicity, we consider two mutually coupled structurally different time-delay systems represented by
\begin{subequations}
\begin{eqnarray}
\vect{\dot{x}}&=&\vect{f}(\vect{x},~\vect{x}_\tau,~\vect{u}), ~~\text{and}\\
\vect{\dot{y}}&=&\vect{g}(\vect{y},~\vect{y}_\tau,~\vect{v}),\qquad \vect{f}\ne \vect{g}
\end{eqnarray}
\label{eqn2a}
\end{subequations}
where $\vect{x}\in\vect{R}^n,~\vect{x}_\tau \in C(\vect{R}^n),~\vect{y}\in\vect{R}^m,~\vect{y}_\tau\in C(\vect{R}^m)$,~$\tau\in\vect{R}$ ($\vect{x}_\tau=\vect{x(t-\tau)},~\vect{y}_\tau=\vect{y(t-\tau)}$) and  $\vect{u},~\vect{v}\in\vect{R}^k,~k\le m,~n$. $u_i=-v_i=h_i\left(\vect{x}(t,\vect{x}_0),~\vect{y}(t,\vect{y}_0)\right)$ correspond to the driving signals. In Eq.~(\ref{eqn2a}), the functions 
$\vect{f}$ and $\vect{g}$ are continuously differentiable or even  continuous functions,
and that their forms are such that there exists only well defined and bounded solution for it.
 System (\ref{eqn2a}) is in GS if there exists a transformation $\vect{H}$ such that the trajectories of the systems (\ref{eqn2a}a) and (\ref{eqn2a}b) are mapped onto a subspace (synchronization manifold) of the whole state space of Eq.~(\ref{eqn2a}). 
%That is
%
%\begin{equation}
%\vect{H}:(\vect{R}^n, \vect{R}^m)\to\vect{R}^{\mu}\subset\vect{R}^n\times \vect{R}^m.
%\end{equation}
%
We also note here that since we are dealing with GS of nonidentical systems with different fractal dimensions, the transformation function $\vect{H}$ refers to a generalized transformation (not the identity transformation) and also there may exist a set of transformations $\vect{H}$ that maps a given $\vect{x},~\vect{x}_{\tau}$ and $\vect{y},~\vect{y}_{\tau}$ to different subspaces of Eq.~(\ref{eqn2a}) \cite{kocarev96}. This indicates that the synchronization manifold $M=\{(\vect{x},~\vect{y}): \vect{H}(\vect{x},~\vect{y})=0\}$ is such that all the initial conditions $\vect{x}(\hat\tau), \vect{y}(\hat\tau), \hat\tau\in\left[-\tau,0\right]$, which lie within a subset of the basin of attraction $B=B_{\vect{x_{\hat\tau}}}\times B_{\vect{y_{\hat\tau}}}$  of Eq.~(\ref{eqn2a}), approaches $M\subset B$ so that $M$ is an attracting manifold. Here $B_{\vect{x_{\hat\tau}}}$ and $B_{\vect{y_{\hat\tau}}}$ are the basins of attraction of systems (\ref{eqn2a}a) and (\ref{eqn2a}b), respectively. The synchronization manifold $M$ can also be
\begin{equation}
M=\{(\vect{x},~\vect{y}): \vect{y}=\vect{H}(\vect{x}) \}~\text{or}~M=\{(\vect{x},~\vect{y}): \vect{x}=\vect{H}(\vect{y})\}
\end{equation}
as special cases, but without ambiguity $M=\vect{H}(\vect{x},~\vect{y})$ is the most general one for mutually coupled systems. Hence, GS exists between the systems (\ref{eqn2a}a) and (\ref{eqn2a}b) only when both coupled systems are asymptotically stable. That is, $\forall~(\vect{x}_i(\hat\tau),~\vect{y}_i(\hat\tau)),~\hat\tau\in\left[-\tau,0\right]\subset B,~i=1,2$, one requires \cite{kocarev96}
\begin{subequations}
\begin{eqnarray}
\lim_{t \to \infty}\vert\vert \vect{y}(t,\vect{x}_1(\hat\tau),~\vect{y}_1(\hat\tau)) -\vect{y}(t,\vect{x}_1(\hat\tau),~\vect{y}_2(\hat\tau))\vert\vert&=&0,\\
\lim_{t \to \infty}\vert\vert \vect{x}(t,\vect{x}_1(\hat\tau),~\vect{y}_1(\hat\tau)) -\vect{x}(t,\vect{x}_2(\hat\tau),~\vect{y}_1(\hat\tau))\vert\vert&=&0.
\end{eqnarray}
\end{subequations}
%
 %This is the mathematical formulation of the auxiliary system approach of the mutually coupled structurally different systems with delay.
Lyapunov stability of an equilibrium means that solutions starting ``close enough" 
to the equilibrium (within a distance $\delta$ from it) remain ``close enough" forever 
(within a distance $\varepsilon$ from it). Note that this must be true for any $\varepsilon$ that one may want to choose.
Asymptotic stability means that solutions that start close enough not only remains close enough but also eventually converge to the equilibrium.  Thus, the asymptotic stability implies that it is Lyapunov stable and there exists $\delta > 0$ such that if $\|x(0)-x_e \|< \delta$, then $\lim_{t \rightarrow \infty} \|x(t)-x_e\| = 0$ \cite{Liapunov1966}.
We note here that the possibility of multi-valued GS occurring in our case is excluded because the main and the corresponding auxiliary systems are starting from different initial conditions in the same basin of attraction. Further, it is worth to emphasize that a subharmonic entrainment takes place when there exists a relation between the interacting systems, which usually takes place for periodic synchronization with $m:n$ periods, $m\neq n$ \cite{parlitz97}. But in this paper, the synchronization dynamics in all the cases we have considered exhibits chaotic/hyperchaotic oscillations and hence the transformation function $\vect{H}$ refers to the existence of a function (not a relation) in our case. Therefore the trajectories of Eq.~(\ref{eqn2a}) starting from the basin of attraction $B$ asymptotically reach the synchronization manifold $M$ defined by the transformation function $\vect{H}(\vect{x},\vect{y})$, which can be smooth if the systems (\ref{eqn2a}) uniformly converge to the GS manifold (otherwise nonsmooth). The uniform convergence (smooth transformation) is confirmed by negative values of their local Lyapunov exponents of the synchronization manifold $M$ \cite{pecora97}.
%
%\begin{figure}
%\centering
%\includegraphics[width=1.0\columnwidth]{sch_diagram.eps}
%\caption{\label{fig1a}Schematic diagrams of the auxiliary system approach for networks of mutually coupled systems. (a, b, d) Linear arrays with $N=2,~3,~4$, (c, e) ring configurations with $N=3,~4$, and (f, g) global and star coupling configurations with $N=4$.}
%\end{figure}
%

Now, we will demonstrate the existence of a transition from partial GS to global GS in symmetrically coupled arbitrary networks of structurally different time-delay systems with different fractal dimensions using the auxiliary system approach. We consider a symmetrically coupled arbitrary network as given in Eq.~(\ref{eqn1}). To determine the asymptotic stability of each of the nodes in this network, one can define a network (auxiliary) identical to Eq.~(\ref{eqn1}) (starting from different initial conditions in the same basin of attraction), whose node dynamics is represented as
\begin{equation}
\vect{\dot{x}}^\prime_i= -\alpha_i \vect{x}^\prime_i(t)+\beta_i \vect{f}_{i}(\vect{x}_i^\prime(t-\tau_{i}))-\varepsilon \sum_{j=1}^N G_{ij}(\vect{x}_{j}-\delta_{ij}\vect{x}^{\prime}_{j}).
\label{eqn2}
\end{equation}
The parameter values are the same as in Eq.(\ref{eqn1}) discussed in Sec.~\ref{sec2}. In the following sections, we will numerically investigate the existence of transition from partial GS to global GS in $N=2$ and $4$ systems with a linear array coupling configurations. To obtain the numerical solution of time-delay systems, it is necessary to convert the continuous evolution of an infinite-dimensional system by a finite number of elements
whose values change at discrete time steps. Hence to calculate the solution $x(t)$ of a delay differential equation for times greater than $t$, a function $x(t)$ over the interval ($t, t-\tau$) must be given. This function can be optimally chosen by $n$ samples taken at intervals $\Delta t=\frac{\tau}{n-1}$. These $n$ samples can equivalently be thought of as the $n$ variables of an $n$-dimensional discrete mapping \cite{lakshmanan10}. In this way a continuous infinite dimensional dynamical system is replaced by a finite, but large, dimensional iterated map.

\section{\label{sec3a}Transition from partial to global GS in $N=2$ mutually coupled time-delay systems}
To start with, we consider a linear array of $N=2$ mutually coupled structurally different time-delay systems. The state equations can be represented as
\begin{subequations}
\begin{eqnarray}
\dot{x}_1&=&-\alpha_1 x_1(t)+\beta_1 f_{1}(x_1(t-\tau_{1}))+\varepsilon (x_{2}-x_{1}),\\
\dot{x}_2&=&-\alpha_2 x_2(t)+\beta_2 f_{2}(x_2(t-\tau_{2}))+\varepsilon (x_{1}-x_{2}).
\end{eqnarray}
\label{2coupa}
\end{subequations}
%
%Now, let us couple the auxiliary system 1$^\prime$ to 2 and 2$^\prime$ to 1 unidirectionally as given in Fig.~\ref{fig1a}(a), such that the systems 1 and 1$^\prime$ are driven by the same signal from system 2, and systems 2 and 2$^\prime$ are similarly driven by system 1. 
The corresponding dynamical equation for the auxiliary systems can be given as
\begin{subequations}
\begin{eqnarray}
\dot{x}^\prime_1&=&-\alpha_1 x_1^\prime(t)+\beta_1 f_{1}(x_1^\prime(t-\tau_{1}))+\varepsilon (x_{2}-x_{1}^\prime),\\
\dot{x}^\prime_2&=&-\alpha_2 x_2^\prime(t)+\beta_2 f_{2}(x_2^\prime(t-\tau_{2}))+\varepsilon (x_{1}-x_{2}^\prime).
\end{eqnarray}
\label{2coupb}
\end{subequations}
We choose the MG time-delay systems (system $x_1$ and $x_1^\prime$) with the nonlinear function $f_{1}(x)$ given in Eq.~(\ref{eqn1a}) and the PWL systems (system $x_2$ and $x_2^\prime$) with the nonlinear function $f_{2}(x)$ given in Eq.~(\ref{eqn1b}). The parameters of both systems are fixed as given in Sec.~\ref{sec2} (Table \ref{table1}) and for those parameter values both systems exhibit hyperchaotic attractors [Figs.~\ref{fig1}(a) and \ref{fig1}(b)] with two [Fig.~\ref{fig2}(a)] and three [Fig.~\ref{fig2}(b)] positive LEs, respectively.

Generally, in a mutual coupling configuration the systems affect each other and attain a common synchronization manifold simultaneously above a threshold value of the coupling strength $\varepsilon$. But interestingly in structurally different coupled time-delay systems with different fractal dimensions, one of the systems first reaches the GS manifold for a lower value of $\varepsilon$, while the other one remains in a desynchronized state, which we call as a partial GS state. For a further increase in $\varepsilon$ both systems organize themselves and reach a common GS manifold, thereby achieving a global GS. In other words, when system $x_1$ and $x_1^\prime$ are identically synchronized, system $x_1$ is synchronized to a subspace (synchronization manifold) 
of the whole state space of both systems in a generalized sense, which we call as a partial GS. Similarly, when the systems $x_2$ and $x_2^\prime$ are synchronized identically, then system $x_2$ is synchronized to the common synchronization manifold. This corroborates that both systems $x_1$ and $x_2$ share a common GS manifold. Thus, when both auxiliary systems are completely synchronized with their original systems for an appropriate coupling strength, then there exists a function that maps systems $x_1$ and $x_2$ to the common (global) GS manifold.
\begin{figure}
\centering
\includegraphics[width=0.8\columnwidth]{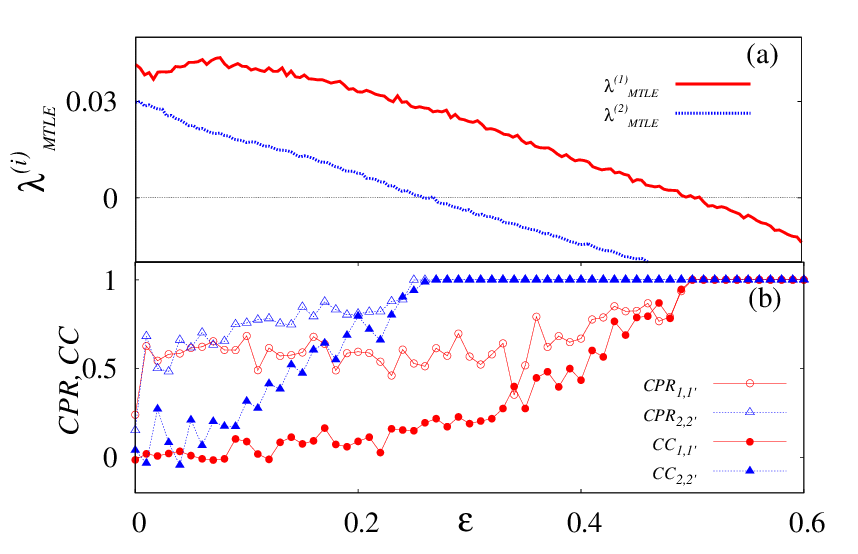}
\caption{\label{fig5} (a) MTLEs and (b) CC, CPR of the main and auxiliary systems for two mutually coupled MG-PWL systems as a function of $\varepsilon\in(0.0,0.6)$ for $N=2$ (\ref{2coupa} and \ref{2coupb}).}
\end{figure}

In order to characterize the transition from partial to global GS and to evaluate the stability of the CS of each of the main and auxiliary systems, we have calculated the MTLEs of the main and auxiliary systems which in turn ensure the stability of GS manifold between the original systems. We have also estimated the correlation coefficient (CC) of each of the main and the associated auxiliary systems, given by
\begin{equation}
C_{i,i^{\prime}} = \frac{\langle(x_{i}(t)-\langle x_{i}(t)\rangle)(x^{\prime}_{i}(t)-\langle x^{\prime}_{i}(t)\rangle)\rangle}{\sqrt{\langle(x_{i}(t)-\langle x_{i}(t)\rangle)^{2}\rangle \langle(x^{\prime}_{i}(t)-\langle x^{\prime}_{i}(t)\rangle)^{2}\rangle}},
\label{eqn5}
\end{equation}
where the $\langle...\rangle$ brackets indicate temporal average. If the two systems are in CS state, the correlation coefficient $CC\approx1$, otherwise $CC<1$. Further, the existence of PS (between the main and auxiliary systems) can be characterized by the value of the index CPR  which can be defined as 
\begin{equation}
CPR=\langle \bar{P_1}(t)\bar{P_2}(t)\rangle/\sigma_1\sigma_2,
\end{equation}
where $P(t)$ is the recurrence-based generalized autocorrelation function defined as
\begin{equation}
P(t)=\frac{1}{N-t} \sum_{i=1}^{N-t} \Theta(\epsilon-||X_i-X_{i+t}|| ).
\end{equation}
Here $\Theta$ is the Heaviside function, $X_i$ is the $i^{th}$ data point of
the system $X$, $\epsilon$ is a predefined threshold, $|| . ||$ is the Euclidean 
norm, and $N$ is the number of data points, $\bar{P}_{1,2}$ means that the 
mean value has been subtracted and
$\sigma_{1,2}$ are the standard deviations of $P_1(t)$ and $P_2(t)$,
respectively. CPR is a recurrence quantification tool mainly used to characterize the phase synchronization in highly non-phase coherent hyperchaotic attractors \cite{marwan07}. If the phases of the coupled systems are mutually locked, then the probability of recurrence is maximal at a time $t$ and CPR $\approx 1$, otherwise the maxima do not occur simultaneously and hence one can expect a drift in both the probability
of recurrence resulting in low values of CPR~\cite{marwan07,senthil06}.

The coupled equations~(\ref{2coupa}) and (\ref{2coupb}), with the nonlinear functions $f_{1}$ and $f_{2}$ as given in equations (\ref{eqn1a}) and (\ref{eqn1b}), respectively, are integrated using a Runge-Kutta fourth order method. The MTLEs are the largest Lyapunov exponents of the evolution equation of $\dot{\Delta}_{i}\equiv\dot{x}_{i}-\dot{x}^{\prime}_{i},~i=1,2$.  The Lyapunov exponents are calculated using J. D. Farmer's approach \cite{lakshmanan10}. In Fig.~\ref{fig5}, we have plotted the various characterizing quantities based on our numerical analysis. The red (light gray) continuous line in Fig.~\ref{fig5}(a) shows the MTLE ($\lambda^{(1)}_{MTLE}$) of the MG systems $(x_{1}, x_{1}^{\prime})$ and the blue (dark gray) dotted line depicts the MTLE ($\lambda^{(2)}_{MTLE}$) of the PWL systems $(x_{2}, x_{2}^{\prime})$ as a function of the coupling strength. Figure \ref{fig5}(b) shows the CC and CPR of the main and auxiliary MG time-delay systems as red (light gray) filled and open circles, respectively, and the CC and CPR of the PWL systems are represented by the blue (dark gray) filled and open triangles. Initially, for $\varepsilon=0$, both $CC_{1,1^{\prime}}$ and $CC_{2,2^{\prime}}$ are nearly zero, indicating the desynchronized state when both $\lambda^{(1)}_{MTLE}$ and $\lambda^{(2)}_{MTLE}>0$ which confirm that CS (GS) is unstable. If we increase the coupling strength, $CC_{2,2^{\prime}}$ and $CPR_{2,2^{\prime}}$ start to increase towards unity and at $\varepsilon^{(2)}_{c}\approx 0.26$, $CC_{2,2^{\prime}}=1$ ($CPR_{2,2^{\prime}}=1$), where $\lambda^{(2)}_{MTLE}<0$, which confirm the simultaneous existence of GS and PS in the PWL system, while the MG system continues to remain in a desynchronized state ($CC_{2,2^{\prime}}\approx 0.2$ and $\lambda^{(1)}_{MTLE}>0$) (partial GS). Further, if we increase the coupling strength to a threshold value $\varepsilon^{(1)}_{c}\approx 0.5$, a global GS occurs where both $CC_{1,1^{\prime}}$ and $CC_{2,2^{\prime}}$ become unity and $\lambda^{(1)}_{MTLE}$ and $\lambda^{(2)}_{MTLE}$ become negative. The transition of the MTLE of the auxiliary and its original systems from positive to negative values as a function of the coupling strength strongly confirms the existence of an attracting manifold. To be more clear, for the value of the coupling strength in the range of global GS, a negative value of the MTLE assures the convergence of the perturbed trajectories in the synchronization manifold (CS between the main and auxiliary systems and GS manifold between the main systems). The convergence corroborates the attracting nature of the synchronization manifold. Further, normally one may expect that the systems with lower dynamical complexity will converge to the GS manifold first, followed by the system with higher dynamical complexity \cite{zheng02}. But to our surprise, we encounter a contrary behavior, where the PWL system with three positive LEs reaches the GS manifold first (at $\varepsilon^{(2)}_{c}\approx0.26$) and then the MG system (with two positive LEs) converges to the GS manifold at $\varepsilon^{(1)}_{c}\approx0.5$ confirming the existence of a transition from partial to global GS in structurally different time-delay systems.
\begin{figure}
\centering
\includegraphics[width=1.0\columnwidth]{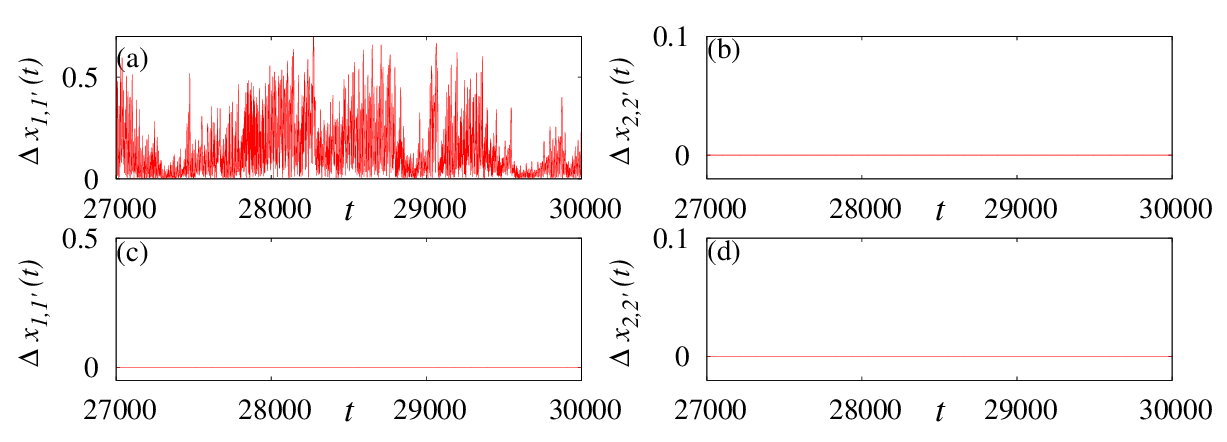}
\caption{\label{fig4}(a, b) The magnitude of difference in the trajectories between the systems ($\Delta x_{i,i^{\prime}}=|x_{i}-x^{\prime}_{i}|, ~~i=1, 2$) for $\varepsilon^{(2)}=0.3$, and (c, d) for $\varepsilon^{(1)}=0.55$ for $N=2$ mutually coupled MG and PWL systems.}
\end{figure}
\begin{figure}
\centering
\includegraphics[width=0.9\columnwidth]{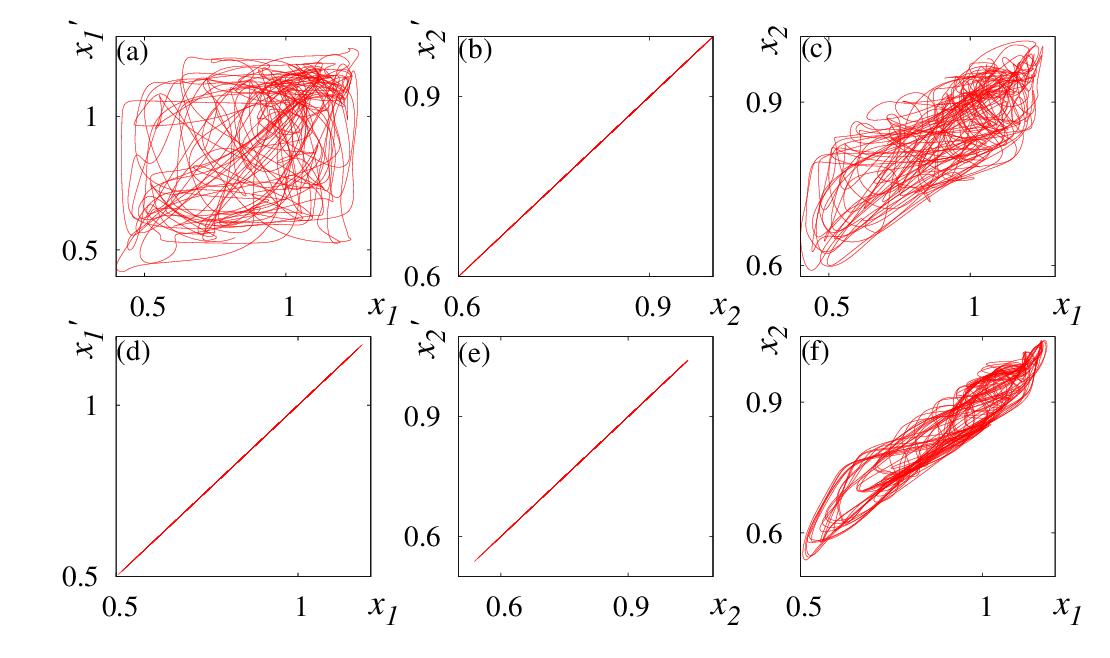}
\caption{\label{fig3}(a-c) The phase portraits of the systems ($x_{1}$, $x^{\prime}_{1}$), ($x_{2}$, $x^{\prime}_{2}$) and ($x_{1}$, $x_{2}$) for $\varepsilon^{(2)}=0.3$, and (d-f) for $\varepsilon^{(1)}=0.55$ for $N=2$.}
\end{figure}

We have also numerically computed the synchronization error ($\Delta x_{i,i^{\prime}}(t)=|x_{i}(t)-x_{i}^{\prime}(t)|, ~~i=1,2$) and phase projection plots, which are depicted in Figs.~\ref{fig4} and \ref{fig3}, respectively. In the absence of the coupling all the systems evolve with their own dynamics. If we slowly increase the coupling strength the main and the auxiliary PWL systems become completely synchronized for $\varepsilon_{c}^{(2)}=0.26$. The synchronization error $\Delta x_{2,2^{\prime}}(t)=0$ and the linear relation between the systems $(x_{2}, x_{2}^{\prime})$ in Fig.~\ref{fig4}(b) and in Fig.~\ref{fig3}(b) (plotted for $\varepsilon^{(2)}=0.3$), respectively, confirm that the systems $x_{2}$ and $x_{2}^{\prime}$ are in a CS state, whereas the MG systems $(x_{1}, x_{1}^{\prime})$ remain desynchronized as confirmed by the phase projection [Fig.~\ref{fig3}(a)] for the same value of the coupling strength. We also note here that for this value of coupling strength the systems $x_{1}$ and $x_{2}$ show certain degree of correlation as depicted in Fig.~\ref{fig3}(c). If we increase $\varepsilon$ further, both sets of systems ($x_{1}, x_{1}^{\prime}$) and ($x_{2}, x_{2}^{\prime}$) reach the CS manifold (for $\varepsilon_{c}^{(1)}=0.5$) and one may expect that both $x_{1}$ and $x_{2}$ attain the common GS manifold, which we call as global GS. Both the synchronization errors $\Delta x_{1,1^{\prime}}(t)$ and $\Delta x_{2,2^{\prime}}(t)$ become zero as shown in Figs.~\ref{fig4}(c) and \ref{fig4}(d) for $\varepsilon^{(1)}=0.55$. This fact confirms the existence of a global GS state. Further, Figs.~\ref{fig3}(d) and \ref{fig3}(e) show a linear relation between the systems ($x_{1}, x^{\prime}_{1}$) and ($x_{2}, x^{\prime}_{2}$), respectively, which additionally confirms the existence of global GS. The degree of correlation in the phase space between the systems $x_{1}$ and $x_{2}$ for the global GS state is depicted in Fig.~\ref{fig3}(f) for the same value of coupling strength.
\begin{figure}
\centering
\includegraphics[width=0.6\columnwidth]{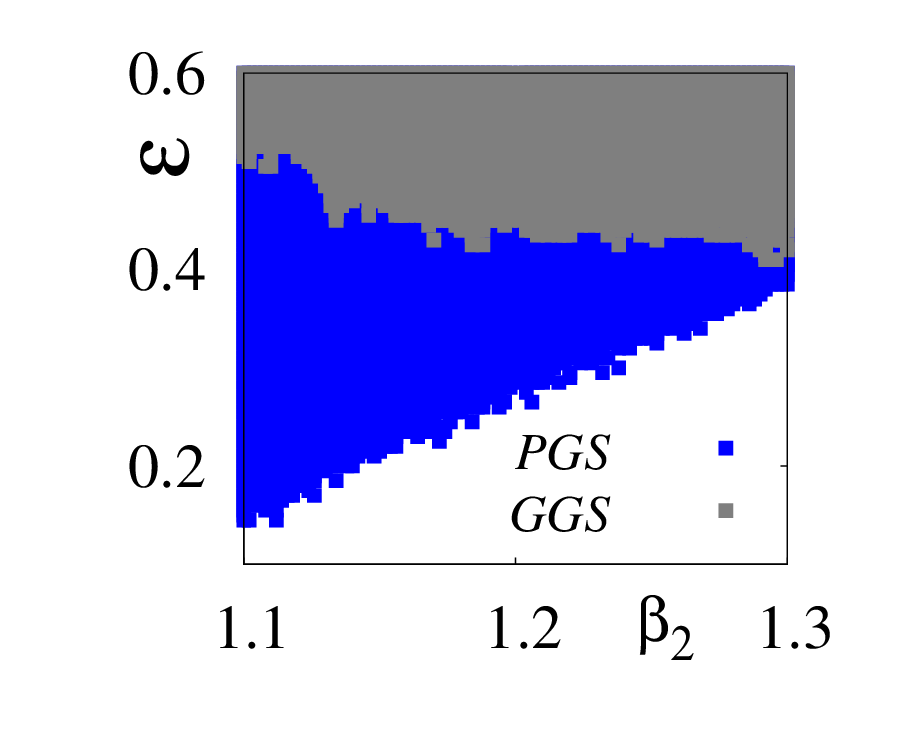}
\caption{\label{fig6}Phase diagram in the ($\beta_{2}-\varepsilon$) plane for two mutually coupled MG-PWL systems showing partial (blue/dark gray), global (light gray) GS and desynchronization (white) regimes.}
\end{figure}

To obtain a global picture on the transition from partial to global GS between the MG and PWL systems, we have plotted the values of $CC_{i,i^{\prime}}$ as a 2-parameter diagram in the ($\beta_{2}-\varepsilon$) plane. We have fixed the parameter values of the MG systems (as given in Sec.~\ref{sec2}) and vary one of the parameter ($\beta_{2}\in(1.1,1.3)$) of the PWL system as a function of the coupling strength ($\varepsilon\in(0.1,0.6)$) as depicted in Fig.~\ref{fig6}. The white region indicates the desynchronized state and the blue (dark gray) region corresponds to the partial GS region, where only one of the mutually coupled systems has reached the common GS manifold as indicated by the unit value of the $CC_{i,i^{\prime}}$. The global GS is represented by light gray where both coupled systems are in GS manifold as confirmed by the unit value of $CC_{i,i^{\prime}}$ of both systems.

We note here that for larger values of the nonlinear parameter $\beta_{2}$ of the PWL system, it needs larger values of coupling strength to attain partial GS whereas global GS is achieved for even smaller coupling strengths than that at lower values of $\beta_2$. This is due to the fact that when we increase the value of $\beta_{2}\in(1.1,1.3)$, the complexity of the PWL system increases and so it requires a larger coupling strength to attain partial GS (blue/dark grey region in Fig.~\ref{fig6}). At the same time, due to the increase in the complexity of the PWL system, it can easily tame the chaotic/hyperchaotic nature of the MG system to reach the common GS manifold even for lower values of $\varepsilon$ than before  (light gray region in Fig.~\ref{fig6}). It is to be emphasized  that during the synchronized state the systems remain in chaotic/hyperchaotic region in the above parameter space.

\section{\label{sec4}Detection of Global GS Using the Mutual False Nearest Neighbor Method}
%The auxiliary system approach have some practical limitations, as this method fails for systems whose dynamical equations are not known (only the measured time series data is available). Even if the system equations are known, the auxiliary response system can be constructed only with finite accuracy (especially, in experimental simulation). Also this method is valid only when the initial conditions of the response and the corresponding auxiliary systems set to be in the same basin of attraction. Due to the above limitations, in this section,
Next, we use the mutual false nearest neighbor method to confirm the existence of global GS in distinctly different time-delay systems. The main idea of this technique consists of the fact of preserving the local neighborliness between the states of the interacting systems \cite{rulkov95}. Let us consider the trajectories of two systems which are connected by the relation $y(t)=\phi (x(t))$ (condition for GS). The MFNN method depends on the observation that in GS state two neighboring points in the phase space of the drive system $x(t)$ correspond to two neighboring points in the phase space of the response system $y(t)$. For mutually coupled systems, the inverse statement is also valid. That is, all close states in the phase space of the system $y(t)$ must correspond to close states of the system $x(t)$.
\begin{figure}
\centering
\includegraphics[width=0.7\columnwidth]{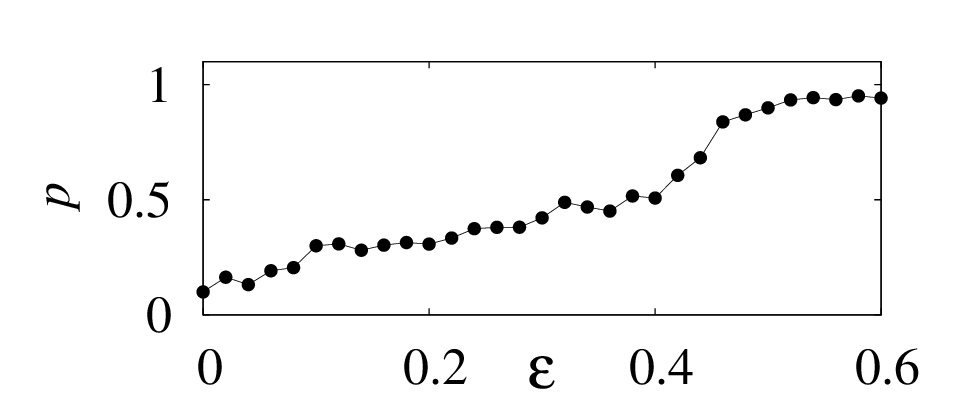}
\caption{\label{fig7}MFNN parameter ($p$) as a function of coupling strength ($\varepsilon$) showing global GS state for $N=2$ mutually coupled MG-PWL time-delay systems (corresponding to Fig.~\ref{fig5}).}
\end{figure}

Let us consider a set of embedded vector points (obtained by attractor reconstruction using time-delay embedding methods \cite{abarbanel93,packard80}) in the spaces of the drive ($X_{1},X_{2},\cdots$) and response ($Y_{1},Y_{2},\cdots$) systems coming from finite segments of the trajectories sampled at uniform intervals of time. Now we can choose an arbitrary point $X_{n}$ in the phase space of the drive system. Let the nearest phase space neighbor of this point in the reconstructed attractor be $X_{nNND}$. In GS state, one can also expect that the corresponding points of the response system $Y_{n}$ will have $Y_{nNND}$ as its close neighbor. From the GS relation, the distance between the two nearest neighbors in the phase space of the response system can be written as $Y_{n}-Y_{nNND}=D\phi (X_{n}) (X_{n}-X_{nNND})$, where $D\phi(X_{n})$ is the Jacobian matrix of the transformation $\phi$ evaluated at $X_{n}$.
%
%\begin{equation}
%Y_{n}-Y_{nNND}=\phi (X_{n})-\phi (X_{nNND}).
%\label{eqn7}
%\end{equation}
%
%As we expect that the difference is small, one can rewrite Eq.~(\ref{eqn7}) as
%
%\begin{equation}
%Y_{n}-Y_{nNND}=D\phi (X_{n}) (X_{n}-X_{nNND}),
%\label{eqn7a}
%\end{equation}
%
%where $D\phi(X_{n})$ is the Jacobian matrix of transformation $\phi$ evaluated at $X_{n}$.
Similarly, we consider the point $Y_{n}$ and locate its nearest neighbor from the time series as $Y_{nNND}$. Again using the GS relation, the distance between the points of the response variables can be written as $Y_{n}-Y_{nNNR}=D\phi (X_{n}) (X_{n}-X_{nNNR})$.
%
%\begin{equation}
%Y_{n}-Y_{nNNR}=D\phi (X_{n}) (X_{n}-X_{nNNR}).
%\label{eqn7b}
%\end{equation}
%
This suggests that the ratio for the MFNN parameter $p$ can be written as,
\begin{equation}
p=\frac{1}{T}\sum_{n}\frac{|Y_{n}-Y_{nNND}|~~|X_{n}-X_{nNNR}|}{|X_{n}-X_{nNND}|~~|Y_{n}-Y_{nNNR}|},
\label{eqn7c}
\end{equation}
where $T$ is the sampling time. In GS state the MFNN parameter $p$ will be of the order of unity. This method has been widely used to identify GS in mutually coupled systems.

% But still we have used MFNN method only in array configuration with $N=2,~3$ and $4$ systems, because in other coupling configurations due to complexity in coupling it is not possible to obtain the reconstructed attractor with similar time-scales in structurally different time-delay systems.

The MFNN parameter ($p$) for two mutually coupled MG and PWL time-delay systems is depicted in Fig.~\ref{fig7} (corresponding to Fig.~\ref{fig5}). As can be seen from this figure the value of $p$ becomes close to unity above $\varepsilon>0.5$, which is indeed the critical coupling strength $\varepsilon^{(1)}_{c}$ in Fig.~\ref{fig5}, strongly confirming the existence of global GS.
\section{\label{sec6}Transition from Partial to Global GS in $N=4$ Mutually Coupled Time-delay Systems}
Further, in this section, we demonstrate the existence of transition from partial to global GS in four (only in $N=4$ for clear visibility of figures depicting synchronization transitions) mutually coupled structurally different time-delay systems in a linear array. In addition to the above two time-delay systems discussed in the previous sections, as the third and fourth systems, we consider the TPWL time-delay system with the nonlinear function given in Eq.~(\ref{eqn1c}) and the Ikeda time-delay system with the nonlinear function as in Eq.~(\ref{eqn1e}). The system parameters are fixed as in
Sec.~\ref{sec2}. For this chosen set of parameter values the TPWL system exhibits a hyperchaotic attractor (Fig.~\ref{fig1}(c)) with four positive LEs ($D_{KY}=8.211$) [see Fig.~\ref{fig2}(c)] and the Ikeda time-delay system exhibits a hyperchaotic attractor [Fig.~\ref{fig1}(d)] with five positive LEs [Fig.~\ref{fig2}(d)] with a KY dimension $D_{KY}=10.116$.
\begin{figure}
\centering
\includegraphics[width=0.8\columnwidth]{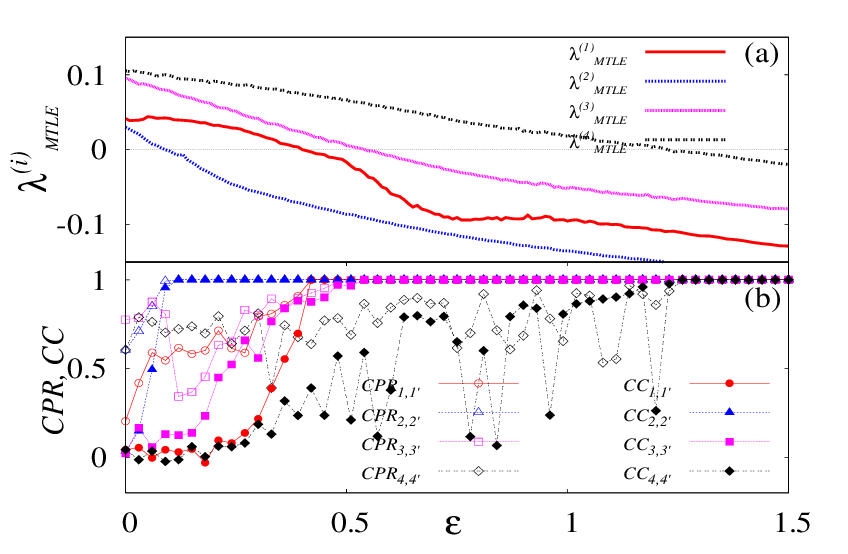}
\caption{\label{fig11}(a) MTLEs and (b) CC, CPR of the main and auxiliary systems for $N=4$ structurally different time-delay systems with linear array configuration as a function of $\varepsilon$.}
\end{figure}
\begin{figure}
\centering
\includegraphics[width=0.8\columnwidth]{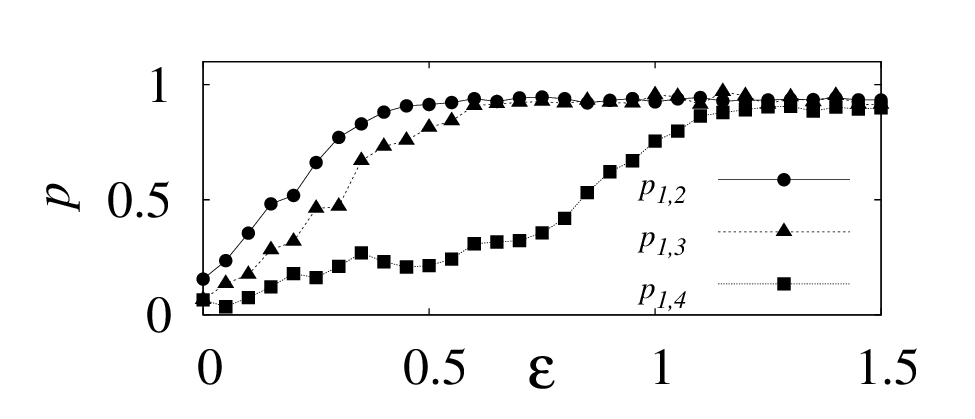}
\caption{\label{fig12}MFNN parameter ($p$) for the linear array of mutually coupled systems with $N=4$ as a function of the coupling strength showing the global GS state (corresponding to Fig.~\ref{fig11}).}
\end{figure}

We demonstrate the existence of a transition from partial to global GS in four mutually coupled time-delay systems (MG, PWL, TPWL and Ikeda) in a linear array configuration. In Figs.~\ref{fig11}(a) and \ref{fig11}(b), we have again presented the characterizing quantities $\lambda^{(i)}_{MTLE}$, $CC_{i,i^{\prime}}$ and $CPR_{i,i^{\prime}}, i,i^{\prime}=1,2,3,4$ of all the four systems along with their associated auxiliary systems as a function of the coupling strength. For $\varepsilon=0$, $\lambda^{(i)}_{MTLE}>0$, while $CC_{i,i^{\prime}}$ and $CPR_{i,i^{\prime}}$ show low values corresponding to a desynchronized state. At $\varepsilon^{(2)}_{c}=0.09$ and $\varepsilon^{(1)}_{c}=0.4$, $\lambda^{(2)}_{MTLE}$ of the PWL system and $\lambda^{(1)}_{MTLE}$ of the MG system become negative, respectively, which confirm that these systems reach the CS (GS) state with their corresponding auxiliary (main) systems, whereas the other two systems are not yet synchronized ($\lambda^{(3,4)}_{MTLE}>0$). Increasing the coupling strength, we find that at $\varepsilon^{(4)}_{c}=0.54$ the Ikeda system with five positive LEs attain the CS (GS) manifold when $\lambda^{(4)}_{MTLE}<0$ and finally the TPWL system becomes synchronized at $\varepsilon^{(3)}_{c}=1.2$, confirming the transition from partial to global GS in four mutually coupled time-delay systems in an array configuration. The $CC_{i,i^{\prime}}$ and $CPR_{i,i^{\prime}}$ of the corresponding systems show a clear transition to unit value for their corresponding threshold values of $\varepsilon^{(i)}_{c}$ confirming the simultaneous existence of GS and PS [Fig.~\ref{fig11}(b)]. 

We have again calculated the MFNN parameter ($p$) for $N=4$ mutually coupled linear array of time-delay systems by considering the MG time-delay system as a reference system (we obtained similar results when we consider any other system as a reference system) as depicted in Fig.~\ref{fig12}. The unit value of the MFNN parameter $p$ of the respective $\varepsilon^{(i)}_{c}$ values of PWL, TPWL and Ikeda time-delay systems (with respect to the MG time-delay system) confirms the existence of the transition from partial to global GS .

Further, we have identified that there exists a similar type of synchronization transition in other regular network configurations like ring, star and global coupling architectures \cite{senthil13} and confirmed that there exits a common GS manifold, where all the systems share to display a global GS, despite their strong heterogeneous nature. In addition, we have also confirmed these synchronization transitions in other permutations on the order of the systems between MG, PWL, TPWL and Ikeda systems in all coupling configurations.  We also wish to emphasize that the synchronization phenomenon examined in our paper is robust against the parameter choice of the dynamical systems.
\section{\label{sec7}Transition from Partial to Global GS in Time-delay Systems of Different Orders}
Now, we will demonstrate the genericity of the transition from partial to global GS by revealing it in coupled time-delay systems of different orders so as to prove that the reported phenomenon is not restricted to time-delay equations with structural similarity alone. In particular, we will show the existence of a transition to global GS via partial GS in such systems using the measures CC and CPR along with the MFNN parameter. We will demonstrate the above results for the following coupled systems.
\begin{enumerate}
\item In a system consisting of a mutually coupled Ikeda time-delay system (which is a scalar first order time-delay system) and a Hopfield neural network~\cite{hopfield82,meng07,lakshmanan10} (which is a second order time-delay system), and
\item In a system of mutually coupled MG time-delay system (which is a scalar first order time-delay system) and a plankton model~\cite{abraham98,gakkhar10} (which correspond to a third order system with multiple delays),
\end{enumerate}

A class of delayed chaotic neural networks~\cite{hopfield82,meng07,lakshmanan10} can be represented as a set of coupled DDEs as given by the equation
\begin{equation}
\dot{x}(t)=-Cx(t)+Af\left[ x(t)\right]+Bf\left[x(t-\tau)\right],
\label{eqn9}
\end{equation}
where $x(t)=\left[ x_1(t),x_2(t),\cdots,x_n(t)\right]^T \in R^n$ is the state vector, the activation function $f\left[ x(t)\right]= \left(f_1\left[ x_1(t)\right],f_2\left[ x_2(t)\right], \cdots,f_n\left[ x_n(t)\right]\right)^T$ denotes the manner in which the neurons respond to each other. $C$ is a positive diagonal matrix, $A=(a_{ij}), i,j=1,2,\cdots,n$ is the feedback matrix, $B=(b_{ij})$ represents the delayed feedback matrix with a constant delay $\tau$. The general class of  delayed neural networks represented by the above Eq.~(\ref{eqn9}) unifies several well known neural networks such as the Hopfield neural networks and cellular neural networks with delay.

The specific set of delayed neural network (Eq.~(\ref{eqn9})) which corresponds to the Hopfield neural network is for the choice of the activation function
\begin{equation}
f\left[ x(t)\right]=\tanh\left[ x(t)\right],
\end{equation}
and for the value of the matrices
\[C=
\begin{bmatrix}
1 &0\\ 0&1
\end{bmatrix},~ A=
\begin{bmatrix}
2.0& -0.1\\ -5.0&3.0
\end{bmatrix},~ B=
\begin{bmatrix}
-1.5& -0.1\\ -0.2&-2.5
\end{bmatrix}.
\]

\begin{figure}
\centering
\includegraphics[width=1.0\columnwidth]{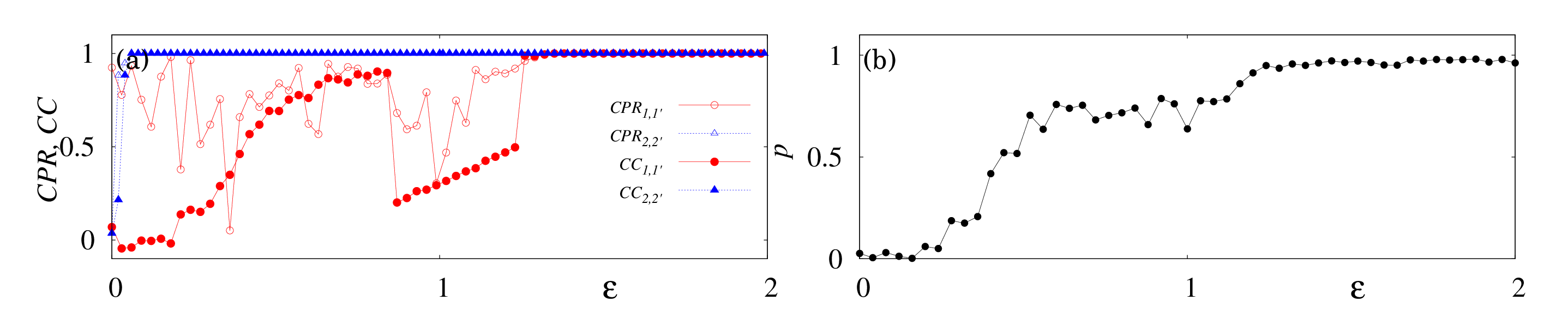}
\caption{\label{fig15}(a) CC and CPR of the main and auxiliary systems and (b) MFNN parameter ($p$) of a coupled Ikeda time-delay system and Hopfield neural network.}
\end{figure}

(i) First we will illustrate the existence of global GS via partial GS in a coupled system consisting of an Ikeda time-delay system, which is mutually coupled to a Hopfield neural network. Mutual coupling is introduced in the $x_{1}(t)$ variable of higher order systems. The CC and CPR between the main and auxiliary systems are depicted in Fig.~\ref{fig15}(a). Low values of CC and CPR in the absence of coupling indicates an asynchronous behavior of both systems. Upon increasing the coupling strength from zero, the second order system, that is the Hopfield neural network, reaches the common synchronization manifold first (partial GS) at the threshold value of $\varepsilon^{(2)}_{c}\approx 0.05$ as indicated by the unit value of the cross correlation coefficient $CC_{2,2^{\prime}}$, while the Ikeda system remains in its transition state (partial GS state).  The simultaneous existence of phase synchronization (PS) together with GS is also confirmed by the unit value of $CPR_{2,2^{\prime}}$ at the same $\varepsilon^{(2)}_{c}$. Further increase in the coupling strength results in the synchronization (both GS and PS) of the Ikeda system to the common synchronization manifold for $\varepsilon^{(1)}_{c}\approx 1.32$ as evidenced from the unit value of $CC_{1,1^{\prime}}$ and $CPR_{1,1^{\prime}}$ [Fig.~\ref{fig15}(a)] confirming the existence of global GS via partial GS in coupled systems of different orders.

We have also calculated the MFNN parameter for the Ikeda system and the Hopfield neural network as a function of the coupling strength, which is depicted in Fig.~\ref{fig15}(b) (corresponding to Fig.~\ref{fig15}(a)). It is evident from this figure that the MFNN parameter $p$ reaches the unit value at $\varepsilon\approx1.32$ perfectly agreeing with the threshold value indicated by CC and CPR confirming the existence of global GS in mutually coupled Ikeda time-delay system and  Hopfield neural network.
\begin{figure}
\centering
\includegraphics[width=1.0\columnwidth]{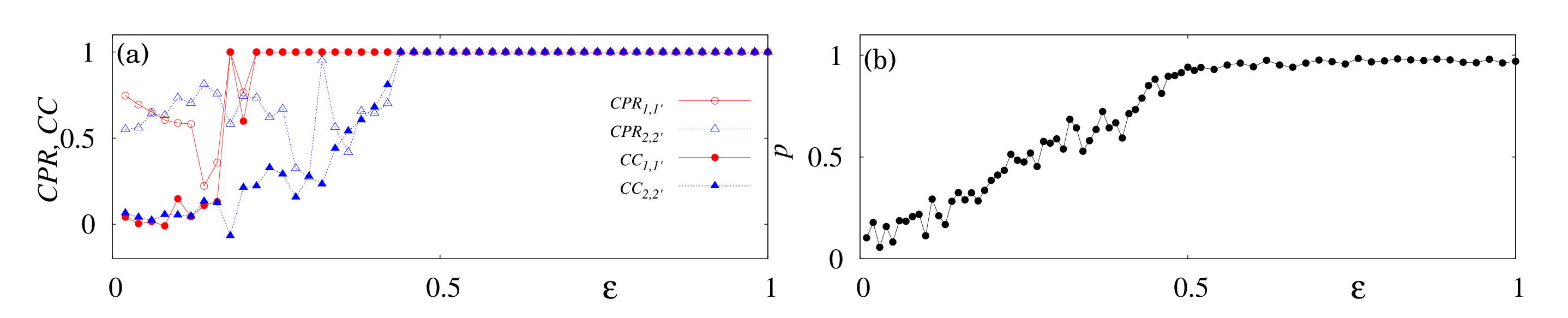}
\caption{\label{fig16}(a) CC and CPR of the main and auxiliary systems and (b) MFNN parameter of a coupled Mackey-Glass time-delay system and the plankton model (\ref{plankton}).}
\end{figure}

(ii) Next, we illustrate the transition from partial to global GS in mutually coupled MG time-delay system, and a third order plankton model~\cite{abraham98,gakkhar10} with multiple delays. The normalized system of equations of a zoo-plankton model is represented as
\begin{subequations}
\begin{align}
\dot{x}=&\,ax[1-(x+y)]-xy-l_1xz,  \\
\dot{y}=&\,xy-b_2y-l_2yz, \\
\dot{z}=&\,-b_1z+l_1x(t-\tau_1)z(t-\tau_3)+l_2y(t-\tau_2)z(t-\tau_3)-n(x+y)z,
\label{plankton}
\end{align}
\end{subequations}
where $a=15.0, b_1=1.0, b_2=0.2, l_1=9.0, l_2=13.5$ and $n=2.0$ are constants. The delays $\tau_{1}, \tau_{2}$ and $\tau_{3}$ are in general different, but for simplicity we have considered identical delays, $\tau_{1}=\tau_{2}=\tau_{3}=5.0$, as studied in Ref \cite{gakkhar10}. Here $x, y$ and $z$ are the normalized quantities of the density of the susceptible phytoplankton, infected phytoplankton and zooplankton (predator species), respectively.  Low values of CC and CPR for $\varepsilon=0$ between the main and auxiliary systems as shown in Fig.~\ref{fig16}(a) confirm that the systems evolve independently in the absence of coupling between them.  As the coupling strength is increased the plankton model synchronizes first to the common synchronization manifold at $\varepsilon^{(1)}_{c}\approx 0.21$ as denoted by $CC_{1,1^{\prime}}=1.0$ indicating partial GS [Fig.~\ref{fig16}(a)]. PS has also occurred simultaneously at the same threshold value of $\varepsilon^{(1)}_{c}$ as indicated by the unit value of $CPR_{1,1^{\prime}}$. Further increase in $\varepsilon$ leads to the existence of global GS by synchronizing the MG time-delay system to the common synchronization manifold as both $CC_{2,2^{\prime}}$ and $CPR_{2,2^{\prime}}$ attain unity at $\varepsilon^{(2)}_{c}\approx 0.44$. The MFNN parameter in Fig.~\ref{fig16}(b) also reaches the unit value at $\varepsilon\approx 0.44$ additionally confirms the occurrence of global GS state.

Hence, it is elucidated that the transition from partial to global GS phenomenon is not restricted to first order structurally different time-delay systems alone but it is also valid for time-delay systems with different orders.
\section{\label{sec8}Conclusion}
In conclusion, we have pointed out the existence of a synchronization transition from partial to global GS in structurally different time-delay systems in symmetrically coupled systems with linear array configuration using the auxiliary system approach and the mutual false nearest neighbor method. We have shown that there exists a smooth transformation function even for networks of structurally different time-delay systems with different fractal dimensions, which maps them to a common GS manifold. We have also found that GS and PS occur simultaneously in structurally different time-delay systems. We have calculated MTLEs to evaluate the asymptotic stability of the CS manifold of each of the main and the corresponding auxiliary systems. This in turn, ensures the stability of the GS manifold between the main systems. In addition, we have estimated the CC and the CPR to characterize the relation between GS and PS. Further, to prove the genericity of our results, we have demonstrated the synchronization transition in systems with different orders such as coupled MG and the Hopfield neural network model and a system of coupled Ikeda and plankton models. We would like to emphasize that now we are working on the experimental realization of the existence of partial and global GS in structurally different time-delay systems using nonlinear time-delayed electronic circuits.
\section*{Acknowledgments}
R. Suresh and D. V. Senthilkumar acknowledges the support from SERB-DST Fast Track scheme for Young Scientists. M. Lakshmanan (M. L.) has been supported by the DST, Government of India sponsored IRHPA research project. M. L. has also been supported by a DAE Raja Ramanna Fellowship.
%
%\end{verbatim}
%\newpage %Just because of unusual number of tables stacked at end
%\bibliography{pre}
%\begin{thebibliography}{42}
%\end{thebibliography}

\end{document}